# Ionic high-pressure form of elemental boron

Artem R. Oganov[1,2]*, Jiuhua Chen[3,4], Carlo Gatti[5], Yanzhang Ma[6], Yanming Ma[1,7], Colin W. Glass[1], Zhenxian Liu[8], Tony Yu[3], Oleksandr O. Kurakevych[9] & Vladimir L. Solozhenko[9]

[1]Laboratory of Crystallography, Department of Materials, ETH Zurich, Wolfgang-Pauli-Str. 10, CH-8093 Zurich, Switzerland. [2]Geology Department, Moscow State University, 119992 Moscow, Russia. [3]Center for the Study of Matter at Extreme Conditions and Department of Mechanical and Materials Engineering, Florida International University, Miami, Florida 33199, USA. [4]Mineral Physics Institute and Department of Geosciences, Stony Brook University, Stony Brook, New York 11794-2100, USA. [5]CNR-ISTM Istituto di Scienze e Tecnologie Molecolari, via Golgi 19, 20133 Milano, Italy. [6]Department of Mechanical Engineering, Texas Tech University, 7th Street & Boston Avenue, Lubbock, Texas 79409, USA. [7]National Laboratory of Superhard Materials, Jilin University, Changchun 130012, China. [8]Geophysical Laboratory, Carnegie Institution of Washington, Washington DC 20015, USA. [9]LPMTM-CNRS, Université Paris Nord, Villetaneuse, F-93430, France.

*Present address: Department of Geosciences and New York Center for Computational Science, Stony Brook University, Stony Brook, New York 11794-2100, USA.

**Boron is an element of fascinating chemical complexity. Controversies have shrouded this element since its discovery was announced in 1808: the new 'element' turned out to be a compound containing less than 60–70% of boron, and it was not until 1909 that 99% pure boron was obtained[1]. And although we now know of at least 16 polymorphs[2], the stable phase of boron is not yet experimentally established even at ambient conditions[3]. Boron's complexities arise from frustration: situated between metals and insulators in the Periodic Table, boron has only three valence electrons, which would favour metallicity, but they are sufficiently localized that insulating states emerge. However, this subtle balance between metallic and insulating states is easily shifted by pressure, temperature and impurities. Here we report the results of high-pressure experiments and *ab initio* evolutionary crystal structure predictions[4,5] that explore the structural stability of boron under pressure and, strikingly, reveal a partially ionic high-pressure boron phase. This new phase is stable between 19 and 89 GPa, can be quenched to ambient conditions, and has a hitherto unknown structure (space group *Pnnm*, 28 atoms in the unit cell) consisting of icosahedral $B_{12}$ clusters and $B_2$ pairs in a NaCl-type arrangement. We find that the ionicity of the phase affects its electronic bandgap, infrared adsorption and**

**dielectric constants, and that it arises from the different electronic properties of the $B_2$ pairs and $B_{12}$ clusters and the resultant charge transfer between them.**

All known structures of boron contain icosahedral $B_{12}$ clusters, with metallic-like three-centre bonds within the icosahedra and covalent two- and three-centre bonds between the icosahedra. Such bonding satisfies the octet rule and produces an insulating state, but impurity-doped boron phases are often metallic. The sensitivity of boron to impurities is evidenced by the existence of unique icosahedral boron-rich compounds such as $YB_{65.9}$, $NaB_{15}$, $MgAlB_{14}$, $AlC_4B_{40}$, $NiB_{50}$ and $PuB_{100}$ (refs 2, 6). In fact, probably only three of the reported boron phases correspond to the pure element[2,7,8]: rhombohedral $\alpha$-$B_{12}$ and $\beta$-$B_{106}$ (with 12 and 106 atoms in the unit cell, respectively) and tetragonal T-192 (with 190–192 atoms per unit cell)[8]. At ambient conditions, $\alpha$-$B_{12}$ and $\beta$-$B_{106}$ have similar static energies[9,10], but disordered $\beta$-$B_{106}$ becomes marginally more stable (in what could seem a violation of the third law of thermodynamics) when zero-point vibrational energy is taken into account[10]. At pressures above several gigapascals, the much denser $\alpha$-$B_{12}$ phase should be more stable at all temperatures. At high pressures, opposing effects come into play: although pressure favours metallic states and might stabilize metallic-like icosahedral clusters[11], the very low packing efficiency of atoms in icosahedral structures (34% for $\alpha$-$B_{12}$) necessitates the destruction of the icosahedra and formation of denser phases (for example, the $\alpha$-Ga-type phase[12]). In experiments, the room-temperature compression of $\beta$-$B_{106}$ showed metastable amorphization[11] at 100 GPa and the onset of superconductivity[13] at 160 GPa. When using laser heating to overcome kinetic barriers, it was found that $\beta$-$B_{106}$ transforms into the T-192 phase above 10 GPa at 2,280 K (ref. 14).

To further explore the intriguing high-pressure behaviour of boron, we have used 99.9999% pure $\beta$-$B_{106}$ to synthesize (both from the melt and from the solid state) about a dozen samples containing dark-grey grains of a hitherto unknown phase of boron (see Methods for details). Single-phase samples were obtained at 12 GPa, 15 GPa and 20 GPa at temperatures of 1,800 K and 2,000 K. At 20 GPa and >2,450 K we observed the formation of the T-192 phase. Electron microprobe and X-ray microprobe analyses show that the new phase does not contain any detectable amounts of impurity atoms, and spatial homogeneity of the samples was confirmed by micro-Raman spectroscopy with a 5 μm beam. The new phase is quenchable down to ambient conditions, where X-ray diffraction patterns were collected and all the peaks indexed with an orthorhombic cell with $a = 5.0544$ Å, $b = 5.6199$ Å and $c = 6.9873$ Å. However, the structure could not be solved with our experimental diffraction data, and was found using the *ab initio* evolutionary algorithm USPEX[4,5] (see Methods, and Supplementary Information sections 1 and 2). USPEX searches for the structure with the lowest theoretical thermodynamic potential without requiring experimental information. However, the use of experimental cell parameter constraints simplifies searches and we took advantage of this. Fully unconstrained searches were performed at higher pressures, where equilibrated experiments have not yet been performed.

We found that the structure of the new phase, which we refer to as $\gamma\text{-}B_{28}$, contains 28 atoms in the unit cell (Fig. 1, Table 1) and belongs to the space group *Pnnm*. This structure has a diffraction pattern very similar to the experimental pattern (Fig. 2). Our calculations show that it is dynamically stable (see Supplementary Information section 3), and thermodynamically more favourable that any other known or hypothetical form of boron between 19 GPa and 89 GPa at 0 K (Fig. 3). In the $\gamma\text{-}B_{28}$ structure, the centres of the $B_{12}$ icosahedra (formed by sites B2–B5; Table 1) form a slightly distorted cubic close packing as in $\alpha\text{-}B_{12}$; but all octahedral voids are occupied by $B_2$ pairs (formed by site B1), and the new phase is thus denser than $\alpha\text{-}B_{12}$. The $\gamma\text{-}B_{28}$ structure resembles a NaCl-type structure, with the $B_{12}$ icosahedra and $B_2$ pairs playing the roles of 'anion' and 'cation', respectively. The average intra-icosahedral bond length is 1.80 Å and the B–B bond length within the $B_2$ pairs is 1.73 Å.

Some metals (for example, Rb and Ba) are known to adopt under pressure broken-symmetry structures, characterized by two sublattices of the same element in different chemical roles[15]. But $\gamma\text{-}B_{28}$ differs in that it is non-metallic, that its two sublattices are occupied not by single atoms but by clusters ($B_{12}$ and $B_2$), and charge transfer between the constituent clusters makes $\gamma\text{-}B_{28}$ a boron boride, $(B_2)^{\delta+}(B_{12})^{\delta-}$. The exact magnitude of charge transfer depends on the definition of atomic charge used, but the results are all qualitatively consistent with each other. A charge estimate based on the differences in the numbers of electrons within equal atom-centred spheres (sphere radii 0.7–1.0 Å) gives $\delta \approx +0.2$, while spherically averaged Born dynamical charges give much higher values, $\delta = +2.2$. Our preferred estimates of charge transfer are based on Bader theory[16], which partitions the total electron density distribution into 'atomic' regions separated by zero-flux (that is, minimum density) surfaces and is thus physically unbiased and ensures maximum additivity and transferability of atomic properties[16]. As listed in Table 1, this approach gives $\delta = +0.34$ in local-orbital (linear combination of atomic orbitals, LCAO) and $\delta = +0.48$ in projector augmented wave (PAW) calculations. (For details of Bader analysis, see Supplementary Information section 4; for details of LCAO basis sets, see Supplementary Information sections 5 and 6.)

Evidence that these large Bader charges originate from the interaction between the $B_2$ and $B_{12}$ clusters comes from the independent atom model (IAM, where the total electron density is a sum of non-interacting atomic densities), which has negligible charges that are an order of magnitude lower than when the atoms are allowed to interact. Similarly, removing the $B_2$ pairs from the $\gamma$-$B_{28}$ structure while maintaining interactions, we again obtain negligible charges (within ±0.03). The calculated electron density distribution directly shows charge transfer giving rise to a strong electron density asymmetry along the B1–B2 line with depletion near B1 and accumulation near B2 (Fig. 4a), which is typical of polar (that is, partially ionic) bonds. The bond asymmetry parameter (see Supplementary Information section 4) for this 'ionic' bond reaches 20%. Because of charge transfer, atomic volumes overall shrink (relative to the IAM) for positively charged and expand for negatively charged atoms (Table 1).

Ionicity affects many properties of $\gamma$-$B_{28}$: it results in large difference between high-frequency ($\varepsilon_\infty$) and static ($\varepsilon_0$) dielectric constants (11.4 and 13.2, respectively), and creates the LO-TO splitting, typical of ionic crystals, and strong infrared absorption. We measured transmission in the far-infrared (100–700 cm$^{-1}$) and mid-infrared (600–1,200 cm$^{-1}$) regions and indeed observed a number of strong absorption bands that compare well the calculated spectrum (Fig. 4b). LO-TO splitting is different for all modes and reciprocal-space directions, and the simplest parameter characterizing it is

$\zeta = \sum_{i=1}^{n} \left( \omega_i^{LO} / \omega_i^{TO} \right)^2 = \varepsilon_0 / \varepsilon_\infty$. In the case of non-ionic crystals, $\zeta = 1$ (we obtain 1.01 for $\alpha$-$B_{12}$); and whenever there is charge transfer, $\zeta > 1$ (we find 1.16 for $\gamma$-$B_{28}$ and 1.18 for GaAs). Among the individual mode splittings in $\gamma$-$B_{28}$, the largest one (337–375 $cm^{-1}$) corresponds to the most intense infrared-active mode, a nearly rigid-body opposite motion of the $B_2$ and $B_{12}$ units.

Although ionic $\gamma$-$B_{28}$ shares structural similarities with covalent $\alpha$-$B_{12}$ (Fig. 1), its electronic structure is quite different: it shows little change of the bandgap with pressure and remains even at 200 GPa an insulator with a relatively wide gap of 1.25 eV (note that density-functional calculations usually underestimate bandgaps by ~40%), whereas the bandgap calculated for $\alpha$-$B_{12}$ rapidly decreases on compression and closes at ~160 GPa (Fig. 4c). These differences in behaviour arise from the presence of charge transfer in $\gamma$-$B_{28}$ and its absence in $\alpha$-$B_{12}$.

The origin and direction of charge transfer in $\gamma$-$B_{28}$ is explained by the electronic properties of the $B_2$ and $B_{12}$ sublattices: the latter is a p-type semiconductor (Fermi level in the valence band) and the former an n-type semiconductor (Fermi level well within the conduction band), so their interaction results in charge transfer from the n-type to the p-type sublattice and the formation of an insulating state for $\gamma$-$B_{28}$ (Fig. 4d). Band-decomposed electron density shows that B2 atoms 'pull' the bonding orbital of $(B1)_2$ to form a polar three-centre two-electron covalent bond B2–B1–B2 (B1–B2 distances are 1.90 Å) and thereby attain a negative charge (Table 1). The top of the valence band (and bottom of the conduction band) in $\gamma$-$B_{28}$ is clearly dominated by the $B_2$ pairs; this, and the spatial separation of the pairs (the nearest $B_2 \ldots B_2$ distance is 3.3 Å), explain why pressure has little effect on the bandgap of $\gamma$-$B_{28}$.

The example of $\gamma$-$B_{28}$ shows that significant ionicity can occur in elemental solids, reinforcing previous theoretical suggestions (for example, hydrogen with $H^+H^-$ molecules[17]). Ionicity appears as a result of many-body interactions, which are strongest under pressure or when atomic orbitals are diffuse (but not diffuse enough to form metallic states)—that is, at the border between metallic and insulating states. This means that systems most favourable for exhibiting ionicity are amphoteric elements close to the Zintl line (B-Si-As-Te-At), and solids made of nanoclusters with very different electronic structures. In the case of boron studied here, our computational and experimental findings show that the ability to form $B_2$ and $B_{12}$ clusters with very different electronic properties is indeed crucial for $\gamma$-$B_{28}$ being ionic. We note that the cationic $B_2^{4+}$ group is well known and its typical B–B distance[6] (1.70–1.75 Å in $B_2F_4$ and $B_2Cl_4$) is the same as in $\gamma$-$B_{28}$ (1.73 Å). The $B_{12}$ cluster is more stable as the $B_{12}^{2-}$ anion (as in the very stable icosahedral $(B_{12}H_{12})^{2-}$ cluster[6]) because its neutral state has an unoccupied bonding orbital[18]. This orbital creates an acceptor band that is located above the valence band edge in boron-rich solids and may be partially occupied by electrons from dopant metal atoms or from other boron clusters (as detected by optical spectroscopy[19]). The $B_2$ pairs in $\gamma$-$B_{28}$ thus behave as electron donors, much like the metal dopants in boron-rich borides.

$\gamma$-$B_{28}$ remains stable up to 89 GPa, when we predict it to transform into the $\alpha$-Ga-type phase (Fig. 3; see also Supplementary Fig. 2), found in our fully unconstrained variable-cell evolutionary simulations at 100 GPa and 300 GPa, in agreement with a previous intuitive proposal[12]. We note that the $\alpha$-Ga-type boron phase, expected to be a poor metal exhibiting superconductivity on cooling[20], is likely to require for its synthesis experiments at >89 GPa and high temperatures (to overcome the activation barrier). The wide stability of insulating states and poor metallicity at higher pressures (at least to >300 GPa) result from the high localization of valence electrons in the boron atom (outermost orbital radius 0.78 Å; compare with 0.62 Å for C, 1.31 Å for Al, 1.25 Å for Ga).

Two centuries after the discovery of boron, we are close to understanding its phase diagram. As illustrated by the schematic phase diagram (Fig. 3 inset), ionic $\gamma$-$B_{28}$ has a large stability field. It achieves its high density and resultant stability under pressure by using 'empty' space to host additional $B_2$ pairs within the $B_{12}$ lattice. Because of the charge transfer between these two components, the new phase can be regarded as a boron boride $(B_2)^{\delta+}(B_{12})^{\delta-}$. Similar ionic interactions could also be important in other pure elemental solids (and probably also liquids) and should give rise to unusual physical properties of fundamental and potentially even practical interest.

## METHODS SUMMARY

All calculations are based on density functional theory within the generalized gradient approximation[21], which gives good results for boron: for $\alpha$-$B_{12}$ at 1 atm, the predicted unit cell parameters are 0.08% smaller than experimental values (Table 1), and bond lengths are within 0.007 Å of experiment. The agreement remains good (<1% errors in cell parameters) also when zero-point vibrational pressure is taken into account.

Evolutionary structure prediction runs were done with the USPEX code[5]; the underlying structure relaxation and total-energy calculations used the frozen-core all-electron PAW method[22] as implemented in the VASP code[23]. Infrared spectra, Born charges and dielectric constants were calculated using density-functional perturbation theory[24] and plane-wave pseudopotential method, as implemented in the ABINIT code[25]. Relaxed-core all-electron LCAO calculations (using the CRYSTAL code[26]) were the main tool for analysing chemical bonding (Bader theory, projected densities of states).

Samples of $\gamma$-$B_{28}$ were synthesized in BN capsules (at high pressures, crystalline boron does not react with BN at temperatures below 2,000 K; ref. 27) using large-volume multianvil apparatuses at Stony Brook, LPMTM-CNRS and at the Geophysical Laboratory, Carnegie Institute of Washington. Conditions of synthesis are: 12 GPa and 1,800 K (annealed for 30 min), 15 GPa and 1,800 K (annealed for 60 min), and 20 GPa and 2,000 K (annealed for 10 min). X-ray diffraction patterns of the recovered samples were collected at beamline X17C of the National Synchrotron Light Source, Brookhaven National Laboratory, at the BW5 beamline of the HASYLAB-DESY, at beamline ID27 of the ESRF, and at the LPMTM-CNRS (TEXT 3000 INEL). For additional details, see Supplementary Information section 7.

[Proofreader: Please update/confirm the tentative publication date]

<edb>1. Petryanov-Sokolov, I. V. (ed.) *Popular Library of the Elements* Ch. 5 (Nauka, 1983).</edb>

<bok>2. Douglas, B. E. & Ho, S.-M. *Structure and Chemistry of Crystalline Solids* (Springer, 2006).</bok>

<bok>3.     Chase, M. W. Jr *J. Phys. Chem. Ref. Data* **9,** (1998).</bok>

<unknown>4. Oganov, A. R. & Glass, C. W. Crystal structure prediction using *ab initio* evolutionary algorithms: Principles and applications. *J. Chem. Phys.* **124,** 244704 (2006).</unknown>

<jrn>5. Glass, C. W., Oganov, A. R. & Hansen, N. USPEX — evolutionary crystal structure prediction. *Comput. Phys. Commun.* **175,** 713–720 (2006).</jrn>

<bok>6.     Wells, A. F. *Structural Inorganic Chemistry* (Clarendon, 1986).</bok>

<jrn>7. Amberger, E. & Ploog, K. Bildung der Gitter des Reinen Bors. *J. Less Common Met.* **23,** 21–31 (1971).</jrn>

<jrn>8. Vlasse, M., Naslain, R., Kasper, J. S. & Ploog, K. Crystal structure of tetragonal boron related to α-AlB$_{12}$. *J. Solid State Chem.* **28,** 289–301 (1979).</jrn>

<unknown>9. Masago, A., Shirai, K. & Katayama-Yoshida, H. Crystal stability of α- and β-boron. *Phys. Rev. B* **73,** 104102 (2006).</unknown>

<jrn>10.     van Setten, M. J., Uijttewaal, M. A., de Wijs, G. A. & de Groot, R. A. Thermodynamic stability of boron: The role of defects and zero point motion. *J. Am. Chem. Soc.* **129,** 2458–2465 (2007).</jrn>

<unknown>11.     Sanz, D. N., Loubeyre, P. & Mezouar, M. Equation of state and pressure induced amorphization of β-boron from X-ray measurements up to 100 GPa. *Phys. Rev. Lett.* **89,** 245501 (2002).</unknown>

<unknown>12.     Häussermann, U., Simak, S.I., Ahuja, R. & Johansson, B. Metal-nonmetal transition in the boron group elements. *Phys. Rev. Lett.* **90,** 065701 (2003).</unknown>

<jrn>13.     Eremets, M. I., Struzhkin, V. W., Mao, H. K. & Hemley, R. J. Superconductivity in boron. *Science* **293,** 272–274 (2001).</jrn>

<unknown>14. Ma, Y. Z., Prewitt, C. T., Zou, G. T., Mao, H. K. & Hemley, R. J. High-pressure high-temperature x-ray diffraction of β-boron to 30 GPa. *Phys. Rev. B* **67,** 174116 (2003).</unknown>

<jrn>15. McMahon, M. I. & Nelmes, R. J. High-pressure structures and phase transformations in elemental metals. *Chem. Soc. Rev.* **35**, 943–963 (2006).</jrn>

<bok>16. Bader, R. F. W. *Atoms in Molecules. A Quantum Theory* (Oxford Univ. Press, 1990).</bok>

<jrn>17. Edwards, B. & Ashcroft, N. W. Spontaneous polarization in dense hydrogen. *Nature* **388**, 652–655 (1997).</jrn>

<jrn>18. Hayami, W. Theoretical study of the stability of $AB_{12}$ (A = H-Ne) icosahedral clusters. *Phys. Rev. B* **60**, 1523–1526 (1999).</jrn>

<jrn>19. Werheit, H., Luax, M. & Kuhlmann, U. Interband and gap state related transitions in β-rhombohedral boron. *Phys. Status Solidi B* **176,** 415–432 (1993).</jrn>

<unknown>20. Ma, Y. M., Tse, J. S., Klug, D. D. & Ahuja, R. Electron-phonon coupling of α-Ga boron. *Phys. Rev. B* **70**, 214107 (2004).</unknown>

<jrn>21. Perdew, J. P., Burke, K. & Ernzerhof, M. Generalized gradient approximation made simple. *Phys. Rev. Lett.* **77**, 3865–3868 (1996).</jrn>

<jrn>22. Blöchl, P. E. Projector augmented-wave method. *Phys. Rev. B* **50**, 17953–17979 (1994).</jrn>

<jrn>23. Kresse, G. & Furthmüller, J. Efficiency of ab initio total-energy calculations for metals and semiconductors using a plane-wave basis set. *Comput. Mater. Sci.* **6,** 15–50 (1996).</jrn>

<jrn>24. Baroni, S., de Gironcoli, S., Dal Corso, A. & Gianozzi, P. Phonons and related crystal properties from density-functional perturbation theory. *Rev. Mod. Phys.* **73**, 515–562 (2001).</jrn>

<jrn>25. Gonze, X. *et al.* First-principles computation of materials properties: The ABINIT software project. *Comput. Mater. Sci.* **25**, 478–492 (2002).</jrn>

<bok>26. Dovesi, R. *et al.* CRYSTAL06 User's Manual (University of Torino, 2006).</bok>

<jrn>27. Solozhenko, V. L., Le Godec, Y. & Kurakevych, O. O. Solid-state synthesis of boron subnitride, $B_6N$: myth or reality? *C.R. Chimie* **9,** 1472–1475 (2006).</jrn>

<jrn>28. Will, G. & Kiefer, B. Electron deformation density in α-boron. *Z. Anorg. Allg. Chem.* **627,** 2100–2104 (2001).</jrn>

<jrn>29. Brazhkin, V. V., Taniguchi, T., Akaishi, M. & Popova, S. V. Fabrication of β-boron by chemical-reaction and melt-quenching methods at high pressures. *J. Mater. Res.* **19,** 1643–1648 (2004).</jrn>

**Supplementary Information** is linked to the online version of the paper at www.nature.com/nature.

**Acknowledgements** A.R.O. acknowledges the Swiss National Science Foundation (grant 200021-111847/1) and the ETH Research Equipment Programme for support of this work. J.C. and T.Y. were supported by the NSF (grant EAR0711321) and the DOE (contract DE-FG02-07ER46461), Yanzhang Ma was funded by the DOE (agreement DE-FC03-03NA00144) and the NSF (grant DMR-0619215), and O.O.K. and V.L.S. were supported by the Agence Nationale de la Recherche (grant ANR-05-BLAN-0141). The use of the NSLS at Brookhaven National Laboratory was supported by the U.S. Department of Energy under Contract DE-AC02-98CH10886, and high pressure beamlines at the NSLS were supported by COMPRES under NSF Cooperative Agreement EAR 06-49658. Calculations were performed at CSCS (Manno), ETH Zurich, and the Joint Supercomputer Centre of the Russian Academy of Sciences.

**Author Contributions** A.R.O. did most of the calculations and wrote most of the paper, C.G. did most of the analysis of chemical bonding and wrote a significant part of the discussion, Yanming Ma contributed to calculations, C.W.G. wrote the first version of the USPEX code, J.C., V.L.S. and Yanzhang Ma synthesized the new phase, J.C. performed infrared absorption measurements with Z.L. and T.Y., V.L.S. did the Raman measurements and elemental analysis, V.L.S. and Yanzhang Ma did synchrotron X-ray diffraction measurements, and J.C. and O.O.K. performed the Le Bail refinement of the X-ray diffraction patterns.

**Author Information** Reprints and permissions information is available at www.nature.com/reprints. Correspondence and requests for materials should be addressed to A.R.O. (artem.oganov@sunysb.edu)

**Figure 1 Structures of α-$B_{12}$ and γ-$B_{28}$. a,** α-$B_{12}$; **b,** γ-$B_{28}$. For the partially ionic γ-$B_{28}$ structure, two oppositely charged sublattices are marked by different colours (anionic, blue; cationic, orange). Unit cell vectors are shown.

**Figure 2 Calculated and measured X-ray diffraction patterns of γ-$B_{28}$.** X-ray wavelength is $\lambda = 0.31851$ Å. The $B_2$ pairs have a strong effect on the intensity of low-

angle ($2\theta < 7°$) peaks and are necessary for reproducing the experimental pattern. Inset, one of the samples (0.4 mm in the longest dimension).

**Figure 3 Stability of boron phases.** Enthalpies are shown relative to α-$B_{12}$. Phase transformations occur at 19 GPa (α-$B_{12}$ to γ-$B_{28}$) and 89 GPa (γ-$B_{28}$ to α-Ga-type). $B_{50}$ is not stable for pure boron, in agreement with experiment[7], whereas γ-$B_{28}$ has a wide stability field. For β-$B_{106}$ we used the structural model from ref. 10. For the disordered T-192 phase, we investigated one model with full occupancy of all sites (196 atoms per cell) and three models with full occupancies of all sites but one (192 atoms per cell). The 192-atom models are energetically more favourable and accurately reproduce the observed[14] lattice parameters of T-192. Zero-point energy differences between the phases (3–7 meV per atom) are small and do not change the topology of the phase diagram. Inset, schematic phase diagram of boron, based on present results and previous experimental[14,27,29] and theoretical[10] studies. Colour-coding indicates covalent (green; α-$B_{12}$, β-$B_{106}$, T-192 phases), ionic (yellow; γ-$B_{28}$) and metallic (blue; α-Ga-type) solids. The phase boundary between γ-$B_{28}$ and α-Ga-type phases is based on the static transition pressure (89 GPa) and a Clapeyron slope of −2.71 MPa $K^{−1}$ calculated using density-functional perturbation theory and the generalized gradient approximation (−2.52 MPa $K^{−1}$ using the local density approximation) with the ABINIT code[25]. In these calculations, the dynamical matrices were calculated on $2 \times 2 \times 2$ and $4 \times 2 \times 3$ grids in the Brillouin zone for γ-$B_{28}$ and α-Ga-type phases, respectively, and interpolated on very dense reciprocal-space meshes. From the resulting phonon spectra we computed the entropy ($S$) of each phase and the high-temperature Clapeyron slope, $dP/dT = \Delta S/\Delta V$. The calculated Clapeyron slope for the α–γ transition is −4.7 MPa $K^{-1}$.

**Figure 4 Chemical bonding in γ-$B_{28}$. a**, Electron density along B1-B2 line, showing polar character of the self-consistent density. IAM, independent atom model. **b**, Infrared spectra. The theoretical spectrum was obtained from mode oscillator strengths computed using density-functional perturbation theory; background was added and peaks were broadened by Gaussians with $\sigma = 2.5–21$ $cm^{−1}$. The agreement between theory and experiment is good, except for the intensity of the peak at 300 $cm^{−1}$. This mode is a sliding motion of $x$–$y$ planes, and the discrepancy could be due to non-random orientation

of crystallites in experimental samples, and/or surface or bulk defects (consistent with the
irregular shape of that peak). Arrow denotes the largest LO-TO splitting (337–375 cm$^{-1}$).
**c**, Pressure dependence of the bandgap of α-B$_{12}$ and γ-B$_{28}$. **d**, Total electronic densities of
states in γ-B$_{28}$ and in its B$_2$ and B$_{12}$ sublattices (in the same configurations as in γ-B$_{28}$),
with Fermi levels indicated by thick vertical lines.

## METHODS

### PAW calculations

We used a PAW potential with a [He] core (radius 1.7 a.u.) together with a well-
converged plane-wave basis set (420 eV kinetic energy cut-off) and dense **k**-point meshes
of 0.1 Å$^{-1}$ resolution for sampling the Brillouin zone. Denser **k**-point meshes (resolution
0.05 Å$^{-1}$ or better) were used for the calculations of enthalpy differences (Fig. 3);
electronic densities of states and bandgaps were explored with extremely dense meshes
($12 \times 12 \times 12$ for γ-B$_{28}$ and $18 \times 18 \times 18$ for α-B$_{12}$). Kohn-Sham equations were solved
iteratively, with a self-consistency threshold of $2 \times 10^{-5}$ eV per cell. Structure relaxation
proceeded until the total energy changes were below $2 \times 10^{-4}$ eV per cell. We have
checked that the cut-off of 420 eV gives excellent convergence for pressure (within
0.1 GPa) and energy differences (within $10^{-4}$ eV per atom). Evolutionary crystal structure
prediction runs were performed using the PAW method, and results presented in Table 1
and Figs 1, 2, 3, 4c have been obtained using PAW calculations.

### Pseudopotential calculations

Computational conditions were similar to the PAW calculations, except the use of a
norm-conserving pseudopotential and the associated higher plane-wave cut-off of 884 eV
to enable similar convergence. Infrared spectra (Fig. 4b), Born charges and dielectric
constants were calculated using density-functional perturbation theory[24] for fully re-
relaxed structures at 1 atm. Preliminary Bader analysis was also performed using
pseudopotential calculations (total electron density was obtained as a sum of core and
valence densities) and yielded practically the same results as the more accurate PAW and
LCAO calculations reported here.

**LCAO calculations**

The Kohn-Sham matrix was diagonalized at an isotropic net of $8 \times 8 \times 8$ **k**-points and the same mesh was used in the reciprocal space integration for Fermi energy calculation and density matrix reconstruction. Details of the basis sets are given in Supplementary Information section 6. Results presented in Table 1 and Fig. 4a, d have been obtained using LCAO calculations. LCAO calculations were the main tool for Bader analysis reported here.

**Comparison between PAW and LCAO calculations**

The main differences between these calculations are in the completeness of the basis set used and in the treatment of core electrons. Well-converged plane-wave (PAW and pseudopotential) calculations use complete basis sets, but treat the core electrons as frozen. Core electrons are considered implicitly in pseudopotential calculations or explicitly in the PAW method. LCAO calculations have a somewhat less complete basis set, but treat core and valence electrons on the same footing, enabling fully variational treatment of the core orbitals (including, for example, core polarization effects). With the valence quadruple-$\zeta$ basis set used here, the consistency between the two approaches is remarkable: the energy difference between $\alpha$-$B_{12}$ and $\gamma$-$B_{28}$ is 0.0271 eV per atom in PAW calculations, and 0.0261 eV per atom in LCAO calculations.

**Experiment**

Given the extreme sensitivity of boron to impurities, we used chemically pure starting material (99.9999% pure $\beta$-$B_{106}$), conducted experiments in BN capsules that do not react with boron at temperatures below 2,000 K (ref. 27), and checked the chemical purity and homogeneity of the recovered samples. The homogeneity of all $\gamma$-$B_{28}$ samples was established by micro-Raman spectroscopy (Dilor XY system, 5 μm beam). X-ray electron probe microanalysis (S400, Leica/PGT Spirit and SX-50 Camebax, Cameca) of the recovered samples revealed that the impurities, if present in the $\gamma$-$B_{28}$ samples, are at an undetectable level.

For more technical information, see the following sections of Supplementary Information: 1, details of evolutionary simulations; 2, illustration of an evolutionary simulation, making icosahedra by evolution; 3, lattice dynamics calculations and dynamical stability of the $\gamma$-$B_{28}$ structure; 4, Bader analysis; 5, derivation of the LCAO basis sets; 6, basis set tables; 7, additional experimental details.

**Table 1 Structures of stable boron phases, optimized at 1 atm.**

| Wyckoff position | $x$ | $y$ | $z$ | $Q_{PAW}$ | $Q_{LCAO\text{-}QZ}$ | $Q_{IAM}$ | $V_{LCAO\text{-}QZ}$ (a.u.) | $V_{IAM}$ (a.u.) |
|---|---|---|---|---|---|---|---|---|
| **$\gamma$-B$_{28}$*** | | | | | | | | |
| B1 (4g) | 0.1702 | 0.5206 | 0 | +0.2418 | +0.1704 | +0.0250 | 47.65 | 49.94 |
| B2 (8h) | 0.1606 | 0.2810 | 0.3743 | −0.1680 | −0.1430 | −0.0153 | 48.46 | 46.25 |
| B3 (8h) | 0.3472 | 0.0924 | 0.2093 | +0.0029 | +0.0218 | +0.0035 | 47.17 | 46.90 |
| B4 (4g) | 0.3520 | 0.2711 | 0 | +0.0636 | +0.0301 | −0.0003 | 44.93 | 46.40 |
| B5 (4g) | 0.1644 | 0.0080 | 0 | +0.0255 | +0.0419 | −0.0011 | 46.38 | 47.60 |
| **$\alpha$-B$_{12}$†** | | | | | | | | |
| B1 (18h) | 0.0103 (0.0102) | 0.0103 (0.0102) | 0.6540 (0.6536) | +0.0565 | +0.0416 | −0.0030 | 47.64 | 49.05 |
| B2 (18h) | 0.2211 (0.2212) | 0.2211 (0.2212) | 0.6305 (0.6306) | −0.0565 | −0.0416 | +0.0030 | 50.43 | 49.01 |
| **$\alpha$-Ga structure‡** | | | | | | | | |
| B1 (8f) | 0 | 0.1558 | 0.0899 | 0 | 0 | 0 | 43.08 | 43.08 |

$Q$, Bader charges; $V$, volumes. Values in parentheses in table and below indicate experimental data (this work, ref. [28]).

*Space group *Pnnm*. $a$ = 5.043 (5.054) Å, $b$ = 5.612 (5.620) Å, $c$ = 6.921 (6.987) Å

†Space group $R\bar{3}$m. $a$ = $b$ = $c$ = 5.051 (5.064) Å, $\alpha$ = $\beta$ = $\gamma$ = 58.04° (58.10°).

‡Space group *Cmca*. $a$ = 2.939 Å, $b$ = 5.330 Å, $c$ = 3.260 Å.